\documentclass[aps,10pt,onecolumn,showpacs,pra,citeautoscript,amsmath,amssymb,floatfix,nofootinbib,superscriptaddress,longbibliography]{revtex4-2}

\usepackage[normalem]{ulem}

\usepackage{amsmath}
\usepackage{amssymb}
\usepackage{bbm}
\usepackage{amsthm, thm-restate}

\usepackage{graphicx}

\usepackage{xcolor}

\usepackage{natbib}
\usepackage{hyperref}
\usepackage[capitalise]{cleveref}
\crefname{figure}{Figure}{figures}

\usepackage{algpseudocode}
\usepackage{enumerate}
\usepackage{enumitem}

\definecolor{dgreen}{rgb}{0,.5,.0}

\newcounter{lemmaN}

\newcounter{lemmaA}

\newcounter{colN}

\makeatletter
\newcommand*{\balancecolsandclearpage}{%
  \close@column@grid
  \clearpage
  \twocolumngrid
}
\makeatother

\makeatletter
\def\tocdepth@fullmunge{%
\let\l@section@saved\l@section
\let\l@section\@gobble@tw@
\let\l@subsection@saved\l@subsection
\let\l@subsection\@gobble@tw@
}%
\def\tocdepth@fullrestore{%
\let\l@section\l@section@saved
\let\l@subsection\l@subsection@saved
}%
\newcommand{\hidetoc}[0]{\addtocontents{toc}{\string\tocdepth@fullmunge}}
\newcommand{\restoretoc}[0]{\addtocontents{toc}{\string\tocdepth@fullrestore}}
\makeatother

\usepackage{mdframed}
\newmdenv[
  backgroundcolor=gray!10,
  linecolor=black,
  linewidth=1pt,
  roundcorner=5pt,
  skipabove=12pt,
  skipbelow=12pt,
  innerleftmargin=10pt,
  innerrightmargin=10pt,
  innertopmargin=10pt,
  innerbottommargin=10pt
]{templatebox}

\usepackage{times} 
\usepackage{setspace} 
\usepackage[margin=1in]{geometry} 

\newcommand{\IQOQI}{Institute for Quantum Optics and Quantum Information,\\ Austrian Academy of Sciences, Boltzmanngasse 3, A-1090 Vienna, Austria}
\newcommand{\Peri}{Perimeter Institute for Theoretical Physics, 31 Caroline Street North, Waterloo, ON N2L 2Y5, Canada}
\newcommand{\VCQ}{Vienna Center for Quantum Science and Technology (VCQ), Faculty of Physics,\\ University of Vienna, Boltzmanngasse 5, A-1090 Vienna, Austria}

\makeatletter
\renewcommand*\l@subsection{\@dottedtocline{1}{1.5em}{2em}}
\makeatother

\begin{document}

\title{Theoretical adversarial collaboration: a template}

\author{Kelvin J. McQueen}
\affiliation{Philosophy Department and Institute for Quantum Studies, Chapman University, Orange, CA 92866, United States}
\author{Markus P.\ M\"uller}
\affiliation{\IQOQI{}}
\affiliation{\VCQ{}}
\affiliation{\Peri{}}

\date{July 15, 2026}

\begin{abstract}
Adversarial collaborations have become increasingly prominent in empirical science, where proponents of competing theories jointly design experiments capable of discriminating between them. Much less attention has been given to theoretical disputes that cannot readily be resolved by experiment. We distinguish theoretical adversarial collaboration from its empirical counterpart and propose a five-step template for conducting it. The procedure is designed to move disputants beyond familiar critique--reply--rejoinder exchanges by requiring sustained clarification, supervised steel-manning of the target theory, and the collaborative development and refinement of objections. Its condition for success is that the collaboration produce insights that substantially exceed the existing literature: either novel, non-obvious arguments show that the theory can meet the objections, or the theory's advocate concedes that new conceptual resources or revisions are required. The template was developed through reflection on, and implemented in, our own adversarial collaboration in philosophy and foundations of physics. We offer it here as a general framework for constructive and publishable progress in philosophy, foundational science, and other areas of inquiry in which disputes cannot readily be resolved by experiment.
\end{abstract}




\maketitle

An adversarial collaboration is a research collaboration that aims to make progress toward resolving disagreement among the collaborators. Adversarial collaborations have been explicitly conducted in science since at least the 1980s \cite{latham1988resolving}. They were championed by Kahneman, who saw them as an effective alternative to the common but inefficient critique--reply--rejoinder format \cite{kahneman2003experiences}. They have since become common in the social sciences, especially psychology \cite{clark2022keep}. Elsewhere, however, we believe they remain underappreciated and underutilized. This is especially true in our own disciplines, philosophy and the foundations of physics. We therefore distinguish between \textit{empirical} and \textit{theoretical} adversarial collaborations. While the former have been extensively explored and implemented, the latter have received far less attention. To help remedy this, we offer a procedure for conducting a theoretical adversarial collaboration. We hope it can serve as a template in domains where disagreement runs deep but cannot readily be resolved by experiment. We developed and implemented this procedure over several years through two linked projects in which each of us critically evaluated a theory developed by the other \cite{mcqueen2026quantum,McQueenMuellerRealism}; here we abstract from the discipline-specific details.

\textit{Empirical} adversarial collaborations are common and involve advocates of competing theories working together to construct feasible experiments that distinguish their theories \cite{mellers2001frequency}. Crucially, disputants try to define experimental outcomes that would falsify, or at least reduce their confidence in their theory \cite{corcoran2023accelerating}. They are therefore more than just ordinary attempts to decide between competing theories, as in the famous Eddington experiment that supported general relativity over classical gravity theories \cite{Dyson1920}. Existing collaborations often have catalysts. In what appears to be the first explicit such collaboration, the two teams had previously ``obtained markedly different results" in their independently conducted experiments \cite{latham1988resolving}. The largest and currently ongoing empirical adversarial collaboration, COGITATE, arose in part because the competing theories ``have evolved independently, without cross-talk" \cite{melloni2023adversarial}. Its first experiment has now been published \cite{cogitate2025adversarial}, while dissemination of the results and data from its second experimental paradigm remains ongoing. For other notable examples of empirical adversarial collaboration, see \cite{bateman2005testing,cowan2020scientific,killingsworth2023income}. Many such collaborations have employed neutral mediators to help ensure objectivity in the design and execution of experiments and the interpretation of their results.

\textit{Theoretical} adversarial collaborations are much less common, since they concern disputes that cannot readily be resolved by experiment. Without such an objective standard, it can be much harder to lead a theorist to revise or relinquish their position, raising the question of what would count as successful progress. Indeed, we have not found a single journal article in philosophy that applies a rigorous adversarial procedure to a purely theoretical philosophical dispute.\footnote{There are, however, important neighboring cases. Fischer, Engelhardt, and Sytsma report an empirical adversarial collaboration in experimental ordinary-language philosophy \cite{fischer2021inappropriate}. Crusius et al.\ conduct an adversarial review of competing theories of envy \cite{crusius2020envy}, while Richter and Steinert apply a theoretical adversarial-collaboration approach to the ethical and legal controversy over coercion in psychiatry \cite{richter2026coercion}. Gefaell and Uller have also recently proposed ways of adapting adversarial collaboration to theoretical and meta-scientific disagreements in ecology and evolutionary biology \cite{gefaell2026rivals}.} This contrasts with empirical adversarial collaborations, for which a dedicated submission format has been introduced in at least one psychology journal \cite{rakow2015rationale}. In philosophy, there are many books that bring disputants into sustained exchange (e.g.\ \cite{hurlburt2007describing,fischer2024four,kind2023consciousness,Shoemaker1984-SHOPIG,dennett2011science,smart1973utilitarianism,glasgow2019race,leal_marraud_2022,Armstrong1996-ARMDAD,dennett2021just}). Some of these come close to genuine adversarial collaboration, but most retain a critique--reply--rejoinder or staged-debate format: the disputants contribute separately authored essays, replies, and rejoinders, or the exchange takes the form of a dialogue resembling a transcript of a live debate. In the kind of theoretical adversarial collaboration we envisage, the disputants instead work together to design a procedure for producing well-defined progress, and the collaboration yields a publishable result only when an agreed condition for progress has been met.

 Central to our procedure is a \textit{condition for success} in a theoretical adversarial collaboration. The condition is met when the collaboration produces substantive insights into theory T that go significantly beyond the existing literature. This can occur in two main ways. First, the advocate of T may concede that the theory's existing conceptual resources are insufficient to address objections that have arisen during the collaboration, so that T must be revised or supplemented. Second, the critic may concede that an objection can be met without revising T, but only through a novel and non-obvious argument based on the theory's existing commitments. In practice, a collaboration may yield a mixed outcome: some objections may expose genuine gaps requiring revisions, while others may reveal previously unrecognized consequences or defenses of the existing theory. Here and throughout, ``theory" is used as shorthand: depending on the dispute, T might instead be an idea, argument, proof, policy proposal, or other contested claim.
 
Our condition for success in theoretical adversarial collaboration is useful for at least two reasons. First, it ensures that the collaboration yields a tangible result that clearly marks progress. Second, it does so without requiring the advocate to abandon T entirely or the critic to withdraw all skepticism---outcomes that are often unrealistic in deeply theoretical disputes. The template therefore applies most naturally where one party defends a theory and another raises objections to it. We formulate the procedure in terms of an \textit{advocate} and a \textit{critic} of some theory T, although either role may be occupied by a team of researchers, rather than an individual.

Our proposed procedure is a 5-step process. In the first step \textbf{(1)}, the disputants work together to clarify the main points of contention. One major goal is to identify the ``weak points" of the theory, or the aspects on which the critic believes critical attention should be focused. Once these points are clear, the disputants can try to agree on what kinds of evidence, argument, or reasoning would count either as adequately answering the critic's concern or as showing that T requires revision.

In the second step \textbf{(2)}, the critic actually steps into the shoes of the advocate, and gets a feel for what it is like to defend T - with an emphasis on the ``weak points" of T, specified in the previous step. That is, the critic ``steelmans" T, by trying to formulate the strongest version of T that the critic can, in the critic's own terms. This process has a number of benefits. First, it helps the critic avoid strawmanning T. That is, it will enable the critic to formulate the strongest version of T before criticizing it. Second, with the advocate supervising this, the critic will acquire a strong understanding of T, and so will be much less likely to formulate objections that are based on misunderstanding. Third, if the critic tries to steelman T in talks or even in discussion with others, the critic may acquire fresh perspectives and discover novel objections to T. 

In the third step \textbf{(3)}, there is now a focused attempt by the critic to formulate objections to T. The development of these objections is ``supervised" by the advocate, to ensure that they are not based on a misunderstanding or incorrect formulation of T. Through debate and discussion, these objections may be continually refined and revised.

In the fourth step \textbf{(4)}, the advocate formulates responses to the objections using T's existing conceptual resources, for example, concepts and arguments developed in existing publications defending T. If a response straightforwardly answers an objection without requiring revisions to T or generating an interesting novel argument, the collaboration has not yet produced a substantive result worth publishing.

If every objection were answered straightforwardly in this way, then step 3 would have failed to generate a substantive challenge, and the collaboration would not yet have met its condition for success. A successful collaboration therefore requires the iterative process of steps 3 and 4 to yield at least one of the two outcomes in step 5. 

The first outcome is \textbf{5a}: the critic concedes that an objection can be met without modifying T, but only through a novel and non-obvious argument grounded in the published formulation of T. The main claims of T need not change, but their consequences must be developed beyond the arguments already in the literature. These novel arguments are the publishable result of the adversarial collaboration.

The second outcome is \textbf{5b}: the advocate concedes that new conceptual resources, revisions, or additions to T are required. Here the advocate acknowledges that T's existing commitments are insufficient to address the critic's objections, even with novel arguments, at least within the time allotted to the collaboration. The revised version of T, or the new ideas introduced to defend it, then becomes the publishable product of the collaboration. In practice, some objections may produce outcome 5a and others outcome 5b.

Our own collaboration developed through two complementary projects in which we exchanged the roles of advocate and critic. In the first, McQueen acted as advocate for an IIT-based consciousness-collapse proposal jointly developed with David Chalmers \cite{chalmers2022consciousness}, while M\"uller pressed the question of whether the proposed dynamics could retain its claimed empirical tractability. Together with Ian Durham, who served as a neutral mediator, we constructed a minimal test case and found that making collapse rates track qualitative differences between conscious states requires a surprisingly rich proliferation of dynamical terms \cite{mcqueen2026quantum}. In the second project, M\"uller acted as advocate for his theory of Algorithmic Idealism, whose technical foundations were developed in \cite{muller2020law}, while McQueen first constructed a supervised steel-man of the theory and then formulated objections concerning its conceptual foundations and its proposed resolution of observer paradoxes \cite{McQueenMuellerRealism}. In both cases, the aim was not to secure a victory for one side, but to identify the strongest version of the view under discussion, isolate its genuine pressure points, and determine whether the available conceptual resources could meet them. The template below systematizes the procedure that emerged from these projects and incorporates our reflections on what proved productive (and what proved less productive) in conducting them.

The procedure is summarized in the following template:

 \newpage

\begin{templatebox}
\begin{spacing}{1.1} 
\begin{center}
    \textbf{\large A Template for Theoretical Adversarial Collaboration}
\end{center}

\textbf{Participants:}
\begin{itemize}
    \item \textbf{The advocate}: Defender of theory T.
    \item \textbf{The critic}: Critic of theory T.
\end{itemize}

\textbf{Procedure:}
\begin{enumerate}
    \item \textbf{Advocate and Critic Clarify Central Points of Contention}
    \begin{itemize}
        \item Identify the core disagreements regarding T.
          \item Agree on what evidence, argument, or reasoning would be sufficient to shift perspectives.
    \end{itemize}
    \item \textbf{Critic Constructs a Steel-man Version of T, Supervised by the Advocate}
    \begin{itemize}
        \item The critic formulates in their own terms the strongest possible version of T.
        \item The critic acquires experience explaining and defending T.
        \item The advocate supervises both formulation and defense to ensure accuracy and completeness.
     \end{itemize}
    \item \textbf{Critic Develops Objections to T, Supervised by the Advocate}
    \begin{itemize}
        \item The critic constructs objections to T.
        \item Objections are reviewed by the advocate to confirm their relevance and grounding in the agreed-upon steel-man version of T.
    \end{itemize}
    \item \textbf{Advocate Responds to Objections Using Existing Conceptual Resources}
    \begin{itemize}
        \item The advocate addresses the objections, in terms of conceptual resources in existing publications of T.
        \item Steps 3 and 4 repeat until one or both of the outcomes in step 5 are reached.

    \end{itemize}

        \item \textbf{Resolution: At least one of the following outcomes is achieved}
    \begin{enumerate}[label=5\alph*.]
        \item \textbf{The critic concedes they have not forced revisions}.
        \begin{itemize}
            \item The critic acknowledges that in the time allotted to the collaboration, their objections do not force fundamental revisions or additions to the main claims of T.
            \item However, if step 3 has been successful, the arguments leading the advocate to successfully defend T must have been novel and non-obvious. They are the publishable product of the collaboration.
        \end{itemize}
        \item \textbf{The advocate concedes that new conceptual resources are required.}
        \begin{itemize}
            \item The advocate acknowledges that existing resources are insufficient to address the critic's objections.
            \item The newly revised version of T, or the newly introduced additions needed to defend it, are then the publishable product of the collaboration.
        \end{itemize}
    \end{enumerate}
         
\end{enumerate}

\textbf{Optional Component (to enhance the process):}
\begin{itemize}
    \item \textbf{Neutral Mediator:} An impartial facilitator to ensure constructive dialogue and resolve procedural disputes.
\end{itemize}
\end{spacing} 
\end{templatebox}

\section*{Acknowledgments}

This research was supported by grant number FQXi-RFP-CPW-2015 from the Foundational Questions Institute (FQxI) and Fetzer Franklin Fund, a donor advised fund of Silicon Valley Community Foundation.

This research was supported in part by Perimeter Institute for Theoretical Physics. Research at Perimeter Institute is supported by the Government of Canada through the Department of Innovation, Science, and Economic Development, and by the Province of Ontario through the Ministry of Colleges and Universities.


\bibliography{bibliography}

 \newcommand{\noop}[1]{}
\begin{thebibliography}{30}%
\makeatletter
\providecommand \@ifxundefined [1]{%
 \@ifx{#1\undefined}
}%
\providecommand \@ifnum [1]{%
 \ifnum #1\expandafter \@firstoftwo
 \else \expandafter \@secondoftwo
 \fi
}%
\providecommand \@ifx [1]{%
 \ifx #1\expandafter \@firstoftwo
 \else \expandafter \@secondoftwo
 \fi
}%
\providecommand \natexlab [1]{#1}%
\providecommand \enquote  [1]{``#1''}%
\providecommand \bibnamefont  [1]{#1}%
\providecommand \bibfnamefont [1]{#1}%
\providecommand \citenamefont [1]{#1}%
\providecommand \href@noop [0]{\@secondoftwo}%
\providecommand \href [0]{\begingroup \@sanitize@url \@href}%
\providecommand \@href[1]{\@@startlink{#1}\@@href}%
\providecommand \@@href[1]{\endgroup#1\@@endlink}%
\providecommand \@sanitize@url [0]{\catcode `\\12\catcode `\$12\catcode `\&12\catcode `\#12\catcode `\^12\catcode `\_12\catcode `\%12\relax}%
\providecommand \@@startlink[1]{}%
\providecommand \@@endlink[0]{}%
\providecommand \url  [0]{\begingroup\@sanitize@url \@url }%
\providecommand \@url [1]{\endgroup\@href {#1}{\urlprefix }}%
\providecommand \urlprefix  [0]{URL }%
\providecommand \Eprint [0]{\href }%
\providecommand \doibase [0]{https://doi.org/}%
\providecommand \selectlanguage [0]{\@gobble}%
\providecommand \bibinfo  [0]{\@secondoftwo}%
\providecommand \bibfield  [0]{\@secondoftwo}%
\providecommand \translation [1]{[#1]}%
\providecommand \BibitemOpen [0]{}%
\providecommand \bibitemStop [0]{}%
\providecommand \bibitemNoStop [0]{.\EOS\space}%
\providecommand \EOS [0]{\spacefactor3000\relax}%
\providecommand \BibitemShut  [1]{\csname bibitem#1\endcsname}%
\let\auto@bib@innerbib\@empty
\bibitem [{\citenamefont {Latham}\ \emph {et~al.}(1988)\citenamefont {Latham}, \citenamefont {Erez},\ and\ \citenamefont {Locke}}]{latham1988resolving}%
  \BibitemOpen
  \bibfield  {author} {\bibinfo {author} {\bibfnamefont {G.~P.}\ \bibnamefont {Latham}}, \bibinfo {author} {\bibfnamefont {M.}~\bibnamefont {Erez}},\ and\ \bibinfo {author} {\bibfnamefont {E.~A.}\ \bibnamefont {Locke}},\ }\bibfield  {title} {\bibinfo {title} {Resolving scientific disputes by the joint design of crucial experiments by the antagonists: Application to the erez--latham dispute regarding participation in goal setting},\ }\href@noop {} {\bibfield  {journal} {\bibinfo  {journal} {Journal of Applied Psychology}\ }\textbf {\bibinfo {volume} {73}},\ \bibinfo {pages} {753} (\bibinfo {year} {1988})}\BibitemShut {NoStop}%
\bibitem [{\citenamefont {Kahneman}(2003)}]{kahneman2003experiences}%
  \BibitemOpen
  \bibfield  {author} {\bibinfo {author} {\bibfnamefont {D.}~\bibnamefont {Kahneman}},\ }\bibfield  {title} {\bibinfo {title} {Experiences of collaborative research.},\ }\href@noop {} {\bibfield  {journal} {\bibinfo  {journal} {American Psychologist}\ }\textbf {\bibinfo {volume} {58}},\ \bibinfo {pages} {723} (\bibinfo {year} {2003})}\BibitemShut {NoStop}%
\bibitem [{\citenamefont {Clark}\ \emph {et~al.}(2022)\citenamefont {Clark}, \citenamefont {Costello}, \citenamefont {Mitchell},\ and\ \citenamefont {Tetlock}}]{clark2022keep}%
  \BibitemOpen
  \bibfield  {author} {\bibinfo {author} {\bibfnamefont {C.~J.}\ \bibnamefont {Clark}}, \bibinfo {author} {\bibfnamefont {T.}~\bibnamefont {Costello}}, \bibinfo {author} {\bibfnamefont {G.}~\bibnamefont {Mitchell}},\ and\ \bibinfo {author} {\bibfnamefont {P.~E.}\ \bibnamefont {Tetlock}},\ }\bibfield  {title} {\bibinfo {title} {Keep your enemies close: Adversarial collaborations will improve behavioral science.},\ }\href@noop {} {\bibfield  {journal} {\bibinfo  {journal} {Journal of Applied Research in Memory and Cognition}\ }\textbf {\bibinfo {volume} {11}},\ \bibinfo {pages} {1} (\bibinfo {year} {2022})}\BibitemShut {NoStop}%
\bibitem [{\citenamefont {McQueen}\ \emph {et~al.}(2026)\citenamefont {McQueen}, \citenamefont {Durham},\ and\ \citenamefont {M{\"u}ller}}]{mcqueen2026quantum}%
  \BibitemOpen
  \bibfield  {author} {\bibinfo {author} {\bibfnamefont {K.~J.}\ \bibnamefont {McQueen}}, \bibinfo {author} {\bibfnamefont {I.~T.}\ \bibnamefont {Durham}},\ and\ \bibinfo {author} {\bibfnamefont {M.~P.}\ \bibnamefont {M{\"u}ller}},\ }\bibfield  {title} {\bibinfo {title} {Quantum superpositions of conscious states in a minimal integrated information model},\ }\href@noop {} {\bibfield  {journal} {\bibinfo  {journal} {Entropy}\ }\textbf {\bibinfo {volume} {28}},\ \bibinfo {pages} {394} (\bibinfo {year} {2026})}\BibitemShut {NoStop}%
\bibitem [{\citenamefont {McQueen}\ and\ \citenamefont {M{\"u}ller}(2026)}]{McQueenMuellerRealism}%
  \BibitemOpen
  \bibfield  {author} {\bibinfo {author} {\bibfnamefont {K.~J.}\ \bibnamefont {McQueen}}\ and\ \bibinfo {author} {\bibfnamefont {M.~P.}\ \bibnamefont {M{\"u}ller}},\ }\bibfield  {title} {\bibinfo {title} {Realism about the external world: An adversarial collaboration}} (\bibinfo {year} {2026}),\ \bibinfo {note} {manuscript}\BibitemShut {NoStop}%
\bibitem [{\citenamefont {Mellers}\ \emph {et~al.}(2001)\citenamefont {Mellers}, \citenamefont {Hertwig},\ and\ \citenamefont {Kahneman}}]{mellers2001frequency}%
  \BibitemOpen
  \bibfield  {author} {\bibinfo {author} {\bibfnamefont {B.}~\bibnamefont {Mellers}}, \bibinfo {author} {\bibfnamefont {R.}~\bibnamefont {Hertwig}},\ and\ \bibinfo {author} {\bibfnamefont {D.}~\bibnamefont {Kahneman}},\ }\bibfield  {title} {\bibinfo {title} {Do frequency representations eliminate conjunction effects? an exercise in adversarial collaboration},\ }\href@noop {} {\bibfield  {journal} {\bibinfo  {journal} {Psychological Science}\ }\textbf {\bibinfo {volume} {12}},\ \bibinfo {pages} {269} (\bibinfo {year} {2001})}\BibitemShut {NoStop}%
\bibitem [{\citenamefont {Corcoran}\ \emph {et~al.}(2023)\citenamefont {Corcoran}, \citenamefont {Hohwy},\ and\ \citenamefont {Friston}}]{corcoran2023accelerating}%
  \BibitemOpen
  \bibfield  {author} {\bibinfo {author} {\bibfnamefont {A.~W.}\ \bibnamefont {Corcoran}}, \bibinfo {author} {\bibfnamefont {J.}~\bibnamefont {Hohwy}},\ and\ \bibinfo {author} {\bibfnamefont {K.~J.}\ \bibnamefont {Friston}},\ }\bibfield  {title} {\bibinfo {title} {Accelerating scientific progress through bayesian adversarial collaboration},\ }\href@noop {} {\bibfield  {journal} {\bibinfo  {journal} {Neuron}\ }\textbf {\bibinfo {volume} {111}},\ \bibinfo {pages} {3505} (\bibinfo {year} {2023})}\BibitemShut {NoStop}%
\bibitem [{\citenamefont {Dyson}\ \emph {et~al.}(1920)\citenamefont {Dyson}, \citenamefont {Eddington},\ and\ \citenamefont {Davidson}}]{Dyson1920}%
  \BibitemOpen
  \bibfield  {author} {\bibinfo {author} {\bibfnamefont {F.~W.}\ \bibnamefont {Dyson}}, \bibinfo {author} {\bibfnamefont {A.~S.}\ \bibnamefont {Eddington}},\ and\ \bibinfo {author} {\bibfnamefont {C.}~\bibnamefont {Davidson}},\ }\bibfield  {title} {\bibinfo {title} {Ix. a determination of the deflection of light by the sun's gravitational field, from observations made at the total eclipse of may 29, 1919},\ }\href@noop {} {\bibfield  {journal} {\bibinfo  {journal} {Philosophical Transactions of the Royal Society of London. Series A, Containing Papers of a Mathematical or Physical Character}\ }\textbf {\bibinfo {volume} {220}},\ \bibinfo {pages} {291} (\bibinfo {year} {1920})}\BibitemShut {NoStop}%
\bibitem [{\citenamefont {Melloni}\ \emph {et~al.}(2023)\citenamefont {Melloni}, \citenamefont {Mudrik}, \citenamefont {Pitts}, \citenamefont {Bendtz}, \citenamefont {Ferrante}, \citenamefont {Gorska}, \citenamefont {Hirschhorn}, \citenamefont {Khalaf}, \citenamefont {Kozma}, \citenamefont {Lepauvre} \emph {et~al.}}]{melloni2023adversarial}%
  \BibitemOpen
  \bibfield  {author} {\bibinfo {author} {\bibfnamefont {L.}~\bibnamefont {Melloni}}, \bibinfo {author} {\bibfnamefont {L.}~\bibnamefont {Mudrik}}, \bibinfo {author} {\bibfnamefont {M.}~\bibnamefont {Pitts}}, \bibinfo {author} {\bibfnamefont {K.}~\bibnamefont {Bendtz}}, \bibinfo {author} {\bibfnamefont {O.}~\bibnamefont {Ferrante}}, \bibinfo {author} {\bibfnamefont {U.}~\bibnamefont {Gorska}}, \bibinfo {author} {\bibfnamefont {R.}~\bibnamefont {Hirschhorn}}, \bibinfo {author} {\bibfnamefont {A.}~\bibnamefont {Khalaf}}, \bibinfo {author} {\bibfnamefont {C.}~\bibnamefont {Kozma}}, \bibinfo {author} {\bibfnamefont {A.}~\bibnamefont {Lepauvre}}, \emph {et~al.},\ }\bibfield  {title} {\bibinfo {title} {An adversarial collaboration protocol for testing contrasting predictions of global neuronal workspace and integrated information theory},\ }\href@noop {} {\bibfield  {journal} {\bibinfo  {journal} {PLoS One}\ }\textbf {\bibinfo {volume} {18}},\ \bibinfo {pages} {e0268577} (\bibinfo {year} {2023})}\BibitemShut {NoStop}%
\bibitem [{\citenamefont {Consortium}\ \emph {et~al.}(2025)\citenamefont {Consortium}, \citenamefont {Ferrante}, \citenamefont {Gorska-Klimowska}, \citenamefont {Henin}, \citenamefont {Hirschhorn}, \citenamefont {Khalaf}, \citenamefont {Lepauvre}, \citenamefont {Liu}, \citenamefont {Richter}, \citenamefont {Vidal} \emph {et~al.}}]{cogitate2025adversarial}%
  \BibitemOpen
  \bibfield  {author} {\bibinfo {author} {\bibfnamefont {C.}~\bibnamefont {Consortium}}, \bibinfo {author} {\bibfnamefont {O.}~\bibnamefont {Ferrante}}, \bibinfo {author} {\bibfnamefont {U.}~\bibnamefont {Gorska-Klimowska}}, \bibinfo {author} {\bibfnamefont {S.}~\bibnamefont {Henin}}, \bibinfo {author} {\bibfnamefont {R.}~\bibnamefont {Hirschhorn}}, \bibinfo {author} {\bibfnamefont {A.}~\bibnamefont {Khalaf}}, \bibinfo {author} {\bibfnamefont {A.}~\bibnamefont {Lepauvre}}, \bibinfo {author} {\bibfnamefont {L.}~\bibnamefont {Liu}}, \bibinfo {author} {\bibfnamefont {D.}~\bibnamefont {Richter}}, \bibinfo {author} {\bibfnamefont {Y.}~\bibnamefont {Vidal}}, \emph {et~al.},\ }\bibfield  {title} {\bibinfo {title} {Adversarial testing of global neuronal workspace and integrated information theories of consciousness},\ }\href@noop {} {\bibfield  {journal} {\bibinfo  {journal} {Nature}\ }\textbf {\bibinfo {volume} {642}},\ \bibinfo {pages} {133} (\bibinfo {year} {2025})}\BibitemShut {NoStop}%
\bibitem [{\citenamefont {Bateman}\ \emph {et~al.}(2005)\citenamefont {Bateman}, \citenamefont {Kahneman}, \citenamefont {Munro}, \citenamefont {Starmer},\ and\ \citenamefont {Sugden}}]{bateman2005testing}%
  \BibitemOpen
  \bibfield  {author} {\bibinfo {author} {\bibfnamefont {I.}~\bibnamefont {Bateman}}, \bibinfo {author} {\bibfnamefont {D.}~\bibnamefont {Kahneman}}, \bibinfo {author} {\bibfnamefont {A.}~\bibnamefont {Munro}}, \bibinfo {author} {\bibfnamefont {C.}~\bibnamefont {Starmer}},\ and\ \bibinfo {author} {\bibfnamefont {R.}~\bibnamefont {Sugden}},\ }\bibfield  {title} {\bibinfo {title} {Testing competing models of loss aversion: An adversarial collaboration},\ }\href@noop {} {\bibfield  {journal} {\bibinfo  {journal} {Journal of Public Economics}\ }\textbf {\bibinfo {volume} {89}},\ \bibinfo {pages} {1561} (\bibinfo {year} {2005})}\BibitemShut {NoStop}%
\bibitem [{\citenamefont {Cowan}\ \emph {et~al.}(2020)\citenamefont {Cowan}, \citenamefont {Belletier}, \citenamefont {Doherty}, \citenamefont {Jaroslawska}, \citenamefont {Rhodes}, \citenamefont {Forsberg}, \citenamefont {Naveh-Benjamin}, \citenamefont {Barrouillet}, \citenamefont {Camos},\ and\ \citenamefont {Logie}}]{cowan2020scientific}%
  \BibitemOpen
  \bibfield  {author} {\bibinfo {author} {\bibfnamefont {N.}~\bibnamefont {Cowan}}, \bibinfo {author} {\bibfnamefont {C.}~\bibnamefont {Belletier}}, \bibinfo {author} {\bibfnamefont {J.~M.}\ \bibnamefont {Doherty}}, \bibinfo {author} {\bibfnamefont {A.~J.}\ \bibnamefont {Jaroslawska}}, \bibinfo {author} {\bibfnamefont {S.}~\bibnamefont {Rhodes}}, \bibinfo {author} {\bibfnamefont {A.}~\bibnamefont {Forsberg}}, \bibinfo {author} {\bibfnamefont {M.}~\bibnamefont {Naveh-Benjamin}}, \bibinfo {author} {\bibfnamefont {P.}~\bibnamefont {Barrouillet}}, \bibinfo {author} {\bibfnamefont {V.}~\bibnamefont {Camos}},\ and\ \bibinfo {author} {\bibfnamefont {R.~H.}\ \bibnamefont {Logie}},\ }\bibfield  {title} {\bibinfo {title} {How do scientific views change? notes from an extended adversarial collaboration},\ }\href@noop {} {\bibfield  {journal} {\bibinfo  {journal} {Perspectives on Psychological Science}\ }\textbf {\bibinfo {volume} {15}},\ \bibinfo {pages} {1011} (\bibinfo {year} {2020})}\BibitemShut {NoStop}%
\bibitem [{\citenamefont {Killingsworth}\ \emph {et~al.}(2023)\citenamefont {Killingsworth}, \citenamefont {Kahneman},\ and\ \citenamefont {Mellers}}]{killingsworth2023income}%
  \BibitemOpen
  \bibfield  {author} {\bibinfo {author} {\bibfnamefont {M.~A.}\ \bibnamefont {Killingsworth}}, \bibinfo {author} {\bibfnamefont {D.}~\bibnamefont {Kahneman}},\ and\ \bibinfo {author} {\bibfnamefont {B.}~\bibnamefont {Mellers}},\ }\bibfield  {title} {\bibinfo {title} {Income and emotional well-being: A conflict resolved},\ }\href@noop {} {\bibfield  {journal} {\bibinfo  {journal} {Proceedings of the National Academy of Sciences}\ }\textbf {\bibinfo {volume} {120}},\ \bibinfo {pages} {e2208661120} (\bibinfo {year} {2023})}\BibitemShut {NoStop}%
\bibitem [{\citenamefont {Fischer}\ \emph {et~al.}(2021)\citenamefont {Fischer}, \citenamefont {Engelhardt},\ and\ \citenamefont {Sytsma}}]{fischer2021inappropriate}%
  \BibitemOpen
  \bibfield  {author} {\bibinfo {author} {\bibfnamefont {E.}~\bibnamefont {Fischer}}, \bibinfo {author} {\bibfnamefont {P.~E.}\ \bibnamefont {Engelhardt}},\ and\ \bibinfo {author} {\bibfnamefont {J.}~\bibnamefont {Sytsma}},\ }\bibfield  {title} {\bibinfo {title} {Inappropriate stereotypical inferences? an adversarial collaboration in experimental ordinary language philosophy},\ }\href@noop {} {\bibfield  {journal} {\bibinfo  {journal} {Synthese}\ }\textbf {\bibinfo {volume} {198}},\ \bibinfo {pages} {10127} (\bibinfo {year} {2021})}\BibitemShut {NoStop}%
\bibitem [{\citenamefont {Crusius}\ \emph {et~al.}(2020)\citenamefont {Crusius}, \citenamefont {Gonzalez}, \citenamefont {Lange},\ and\ \citenamefont {Cohen-Charash}}]{crusius2020envy}%
  \BibitemOpen
  \bibfield  {author} {\bibinfo {author} {\bibfnamefont {J.}~\bibnamefont {Crusius}}, \bibinfo {author} {\bibfnamefont {M.~F.}\ \bibnamefont {Gonzalez}}, \bibinfo {author} {\bibfnamefont {J.}~\bibnamefont {Lange}},\ and\ \bibinfo {author} {\bibfnamefont {Y.}~\bibnamefont {Cohen-Charash}},\ }\bibfield  {title} {\bibinfo {title} {Envy: An adversarial review and comparison of two competing views},\ }\href@noop {} {\bibfield  {journal} {\bibinfo  {journal} {Emotion Review}\ }\textbf {\bibinfo {volume} {12}},\ \bibinfo {pages} {3} (\bibinfo {year} {2020})}\BibitemShut {NoStop}%
\bibitem [{\citenamefont {Richter}\ and\ \citenamefont {Steinert}(2026)}]{richter2026coercion}%
  \BibitemOpen
  \bibfield  {author} {\bibinfo {author} {\bibfnamefont {D.}~\bibnamefont {Richter}}\ and\ \bibinfo {author} {\bibfnamefont {T.}~\bibnamefont {Steinert}},\ }\bibfield  {title} {\bibinfo {title} {Can coercion in psychiatry be justified? a theoretical adversarial collaboration approach},\ }\href@noop {} {\bibfield  {journal} {\bibinfo  {journal} {International Journal of Law and Psychiatry}\ }\textbf {\bibinfo {volume} {105}},\ \bibinfo {pages} {102171} (\bibinfo {year} {2026})}\BibitemShut {NoStop}%
\bibitem [{\citenamefont {Gefaell}\ and\ \citenamefont {Uller}(2026)}]{gefaell2026rivals}%
  \BibitemOpen
  \bibfield  {author} {\bibinfo {author} {\bibfnamefont {J.}~\bibnamefont {Gefaell}}\ and\ \bibinfo {author} {\bibfnamefont {T.}~\bibnamefont {Uller}},\ }\bibfield  {title} {\bibinfo {title} {From rivals to partners: Adversarial collaboration in ecology and evolution},\ }\href@noop {} {\bibfield  {journal} {\bibinfo  {journal} {Trends in Ecology \& Evolution}\ }\textbf {\bibinfo {volume} {41}},\ \bibinfo {pages} {37} (\bibinfo {year} {2026})}\BibitemShut {NoStop}%
\bibitem [{\citenamefont {Rakow}\ \emph {et~al.}(2015)\citenamefont {Rakow}, \citenamefont {Thompson}, \citenamefont {Ball},\ and\ \citenamefont {Markovits}}]{rakow2015rationale}%
  \BibitemOpen
  \bibfield  {author} {\bibinfo {author} {\bibfnamefont {T.}~\bibnamefont {Rakow}}, \bibinfo {author} {\bibfnamefont {V.}~\bibnamefont {Thompson}}, \bibinfo {author} {\bibfnamefont {L.}~\bibnamefont {Ball}},\ and\ \bibinfo {author} {\bibfnamefont {H.}~\bibnamefont {Markovits}},\ }\bibfield  {title} {\bibinfo {title} {Rationale and guidelines for empirical adversarial collaboration: A {Thinking \& Reasoning} initiative},\ }\href@noop {} {\bibfield  {journal} {\bibinfo  {journal} {Thinking \& Reasoning}\ }\textbf {\bibinfo {volume} {21}},\ \bibinfo {pages} {167} (\bibinfo {year} {2015})}\BibitemShut {NoStop}%
\bibitem [{\citenamefont {Hurlburt}\ and\ \citenamefont {Schwitzgebel}(2007)}]{hurlburt2007describing}%
  \BibitemOpen
  \bibfield  {author} {\bibinfo {author} {\bibfnamefont {R.~T.}\ \bibnamefont {Hurlburt}}\ and\ \bibinfo {author} {\bibfnamefont {E.}~\bibnamefont {Schwitzgebel}},\ }\href@noop {} {\emph {\bibinfo {title} {Describing Inner Experience? Proponent Meets Skeptic}}}\ (\bibinfo  {publisher} {MIT Press},\ \bibinfo {address} {Cambridge, MA},\ \bibinfo {year} {2007})\BibitemShut {NoStop}%
\bibitem [{\citenamefont {Fischer}\ \emph {et~al.}(2024)\citenamefont {Fischer}, \citenamefont {Kane}, \citenamefont {Pereboom},\ and\ \citenamefont {Vargas}}]{fischer2024four}%
  \BibitemOpen
  \bibfield  {author} {\bibinfo {author} {\bibfnamefont {J.~M.}\ \bibnamefont {Fischer}}, \bibinfo {author} {\bibfnamefont {R.}~\bibnamefont {Kane}}, \bibinfo {author} {\bibfnamefont {D.}~\bibnamefont {Pereboom}},\ and\ \bibinfo {author} {\bibfnamefont {M.}~\bibnamefont {Vargas}},\ }\href@noop {} {\emph {\bibinfo {title} {Four views on free will}}}\ (\bibinfo  {publisher} {John Wiley \& Sons},\ \bibinfo {year} {2024})\BibitemShut {NoStop}%
\bibitem [{\citenamefont {Kind}\ \emph {et~al.}(2023)\citenamefont {Kind}, \citenamefont {Jackson},\ and\ \citenamefont {Stoljar}}]{kind2023consciousness}%
  \BibitemOpen
  \bibfield  {author} {\bibinfo {author} {\bibfnamefont {A.}~\bibnamefont {Kind}}, \bibinfo {author} {\bibfnamefont {F.}~\bibnamefont {Jackson}},\ and\ \bibinfo {author} {\bibfnamefont {D.}~\bibnamefont {Stoljar}},\ }\href@noop {} {\emph {\bibinfo {title} {What is Consciousness?: A Debate}}}\ (\bibinfo  {publisher} {Routledge},\ \bibinfo {year} {2023})\BibitemShut {NoStop}%
\bibitem [{\citenamefont {Shoemaker}\ and\ \citenamefont {Swinburne}(1984)}]{Shoemaker1984-SHOPIG}%
  \BibitemOpen
  \bibfield  {author} {\bibinfo {author} {\bibfnamefont {S.}~\bibnamefont {Shoemaker}}\ and\ \bibinfo {author} {\bibfnamefont {S.}~\bibnamefont {Swinburne}},\ }\href@noop {} {\emph {\bibinfo {title} {Personal Identity: Great Debates in Philosophy}}},\ edited by\ \bibinfo {editor} {\bibfnamefont {R.}~\bibnamefont {Swinburne}}\ (\bibinfo  {publisher} {Blackwell},\ \bibinfo {address} {Oxford, England},\ \bibinfo {year} {1984})\BibitemShut {NoStop}%
\bibitem [{\citenamefont {Dennett}\ \emph {et~al.}(2011)\citenamefont {Dennett}, \citenamefont {Plantinga}, \citenamefont {Plantinga}, \citenamefont {Philosophe},\ and\ \citenamefont {Plantinga}}]{dennett2011science}%
  \BibitemOpen
  \bibfield  {author} {\bibinfo {author} {\bibfnamefont {D.~C.}\ \bibnamefont {Dennett}}, \bibinfo {author} {\bibfnamefont {A.}~\bibnamefont {Plantinga}}, \bibinfo {author} {\bibfnamefont {A.}~\bibnamefont {Plantinga}}, \bibinfo {author} {\bibfnamefont {E.-U.}\ \bibnamefont {Philosophe}},\ and\ \bibinfo {author} {\bibfnamefont {A.}~\bibnamefont {Plantinga}},\ }\href@noop {} {\emph {\bibinfo {title} {Science and religion: Are they compatible?}}}\ (\bibinfo  {publisher} {Oxford University Press Oxford},\ \bibinfo {year} {2011})\BibitemShut {NoStop}%
\bibitem [{\citenamefont {Smart}\ and\ \citenamefont {Williams}(1973)}]{smart1973utilitarianism}%
  \BibitemOpen
  \bibfield  {author} {\bibinfo {author} {\bibfnamefont {J.~J.~C.}\ \bibnamefont {Smart}}\ and\ \bibinfo {author} {\bibfnamefont {B.}~\bibnamefont {Williams}},\ }\href@noop {} {\emph {\bibinfo {title} {Utilitarianism: For and against}}}\ (\bibinfo  {publisher} {Cambridge University Press},\ \bibinfo {year} {1973})\BibitemShut {NoStop}%
\bibitem [{\citenamefont {Glasgow}\ \emph {et~al.}(2019)\citenamefont {Glasgow}, \citenamefont {Haslanger}, \citenamefont {Jeffers},\ and\ \citenamefont {Spencer}}]{glasgow2019race}%
  \BibitemOpen
  \bibfield  {author} {\bibinfo {author} {\bibfnamefont {J.}~\bibnamefont {Glasgow}}, \bibinfo {author} {\bibfnamefont {S.}~\bibnamefont {Haslanger}}, \bibinfo {author} {\bibfnamefont {C.}~\bibnamefont {Jeffers}},\ and\ \bibinfo {author} {\bibfnamefont {Q.}~\bibnamefont {Spencer}},\ }\href@noop {} {\emph {\bibinfo {title} {What is race?: Four philosophical views}}}\ (\bibinfo  {publisher} {Oxford University Press},\ \bibinfo {year} {2019})\BibitemShut {NoStop}%
\bibitem [{\citenamefont {Leal}\ and\ \citenamefont {Marraud}(2022)}]{leal_marraud_2022}%
  \BibitemOpen
  \bibfield  {author} {\bibinfo {author} {\bibfnamefont {F.}~\bibnamefont {Leal}}\ and\ \bibinfo {author} {\bibfnamefont {H.}~\bibnamefont {Marraud}},\ }\href@noop {} {\emph {\bibinfo {title} {How philosophers argue: An Adversarial Collaboration on the Russell--Copleston Debate}}}\ (\bibinfo  {publisher} {Springer International Publishing},\ \bibinfo {year} {2022})\BibitemShut {NoStop}%
\bibitem [{\citenamefont {Armstrong}(1996)}]{Armstrong1996-ARMDAD}%
  \BibitemOpen
  \bibfield  {author} {\bibinfo {author} {\bibfnamefont {D.~M.}\ \bibnamefont {Armstrong}},\ }\href@noop {} {\emph {\bibinfo {title} {Dispositions: A Debate}}},\ edited by\ \bibinfo {editor} {\bibfnamefont {C.~B.}\ \bibnamefont {Martin}}, \bibinfo {editor} {\bibfnamefont {U.~T.}\ \bibnamefont {Place}},\ and\ \bibinfo {editor} {\bibfnamefont {T.}~\bibnamefont {Crane}}\ (\bibinfo  {publisher} {Routledge},\ \bibinfo {address} {New York},\ \bibinfo {year} {1996})\BibitemShut {NoStop}%
\bibitem [{\citenamefont {Dennett}\ and\ \citenamefont {Caruso}(2021)}]{dennett2021just}%
  \BibitemOpen
  \bibfield  {author} {\bibinfo {author} {\bibfnamefont {D.~C.}\ \bibnamefont {Dennett}}\ and\ \bibinfo {author} {\bibfnamefont {G.~D.}\ \bibnamefont {Caruso}},\ }\href@noop {} {\emph {\bibinfo {title} {Just deserts: Debating free will}}}\ (\bibinfo  {publisher} {John Wiley \& Sons},\ \bibinfo {year} {2021})\BibitemShut {NoStop}%
\bibitem [{\citenamefont {Chalmers}\ and\ \citenamefont {McQueen}(2022)}]{chalmers2022consciousness}%
  \BibitemOpen
  \bibfield  {author} {\bibinfo {author} {\bibfnamefont {D.~J.}\ \bibnamefont {Chalmers}}\ and\ \bibinfo {author} {\bibfnamefont {K.~J.}\ \bibnamefont {McQueen}},\ }\bibfield  {title} {\bibinfo {title} {Consciousness and the collapse of the wave function},\ }in\ \href@noop {} {\emph {\bibinfo {booktitle} {Consciousness and Quantum Mechanics}}},\ \bibinfo {editor} {edited by\ \bibinfo {editor} {\bibfnamefont {S.}~\bibnamefont {Gao}}}\ (\bibinfo  {publisher} {Oxford University Press},\ \bibinfo {address} {Oxford},\ \bibinfo {year} {2022})\ pp.\ \bibinfo {pages} {11--63}\BibitemShut {NoStop}%
\bibitem [{\citenamefont {M{\"u}ller}(2020)}]{muller2020law}%
  \BibitemOpen
  \bibfield  {author} {\bibinfo {author} {\bibfnamefont {M.~P.}\ \bibnamefont {M{\"u}ller}},\ }\bibfield  {title} {\bibinfo {title} {Law without law: from observer states to physics via algorithmic information theory},\ }\href@noop {} {\bibfield  {journal} {\bibinfo  {journal} {Quantum}\ }\textbf {\bibinfo {volume} {4}},\ \bibinfo {pages} {301} (\bibinfo {year} {2020})}\BibitemShut {NoStop}%
\end{thebibliography}%

\end{document}